\documentclass[a4paper,fleqn]{cas-dc}

\usepackage{amsmath,amssymb,amsfonts}
\usepackage{algorithm}
\usepackage{algorithmic}
\usepackage{graphicx}
\usepackage{textcomp}
\usepackage{booktabs}
\usepackage[normalem]{ulem}
\usepackage{url}
\usepackage{multirow}
\usepackage{makecell}

\usepackage[authoryear,longnamesfirst]{natbib}

\def\BibTeX{{\rm B\kern-.05em{\sc i\kern-.025em b}\kern-.08em
    T\kern-.1667em\lower.7ex\hbox{E}\kern-.125emX}}
\newcommand{\IEEEPARstart}[2]{#1#2}
\graphicspath{{./}{sections/}}

\begin{document}
\let\WriteBookmarks\relax
\def\floatpagepagefraction{1}
\def\textpagefraction{.001}

\shorttitle{Cross-Receiver Open-Set Radio Frequency Fingerprinting}
\shortauthors{Yao et~al.}

\title [mode = title]{Cross-Receiver Open-Set Radio Frequency Fingerprinting via Structure-Anchored Joint Adaptation}

\author[1]{Fengchong Yao}
\ead{phoenixly@126.com}

\author[1]{Jianbing Li}
\cormark[1]
\ead{li_jb@126.com}

\author[1]{Qing Liu}
\ead{liuqing8123@163.com}

\author[1]{Kefeng Song}
\ead{annx1990@163.com}

\author[1]{Haitao Li}
\ead{lihaitao_01@163.com}

\author[1]{Song Wang}
\ead{wangsong8190@163.com}

\author[1]{Feixiang Wang}
\ead{1061707424@qq.com}

\affiliation[1]{organization={School of Information Systems Engineering, Information Engineering University},
            city={Zhengzhou},
            postcode={450001},
            country={China}}

\cortext[1]{Corresponding author}

\begin{abstract}
Radio frequency fingerprint identification (RFFI) provides a critical physical-layer security mechanism for dynamic Internet of Things (IoT) and ad hoc networks. However, the decentralized and open nature of these networks imposes two strict deployment criteria: the credential must transfer reliably across physically dispersed, heterogeneous receivers, and it must decisively reject unregistered rogue traffic. Cross-receiver hardware shifts depress the confidence of registered devices and may also place unseen rogue transmitters in high-confidence known regions under naive domain adaptation, increasing false acceptance. To address these risks, we propose CRODA-ST, a joint optimization framework that couples Discriminative Structure Anchoring (DSA) with Rejection-Oriented Alignment (ROA). Within this coupled objective, DSA establishes a stable target-known semantic foundation for shifted registered devices, while ROA regularizes the open-set decision boundaries governing rejection of unseen rogue transmitters. In the canonical WiSig setting, CRODA-ST achieves an open-set classification rate (OSCR) of 0.9580 and a target-domain false positive rate of 0.0469 at a 90\% true positive rate ($\mathrm{FPR}_{90}$). A controllable LoRa simulation provides a complementary diagnostic under synthesized hardware distortions. At the distinct source-calibrated deployment operating point with $\rho=0.80$, CRODA-ST yields a target-unknown false acceptance rate (FAR) of 0.0075 in the evaluated setting.
\end{abstract}

\begin{highlights}
\item CRODA-ST addresses source-calibrated cross-receiver open-set RFFI.
\item Target-known anchoring mitigates receiver-induced feature drift.
\item Receiver-oriented regularization reduces unknown false acceptance.
\item WiSig and LoRa tests validate the recognition-rejection balance.
\end{highlights}


\maketitle

\section{Introduction}
\begin{figure*}[!t]
\centering
\includegraphics[width=0.92\textwidth,trim=3mm 3mm 3mm 3mm,clip]{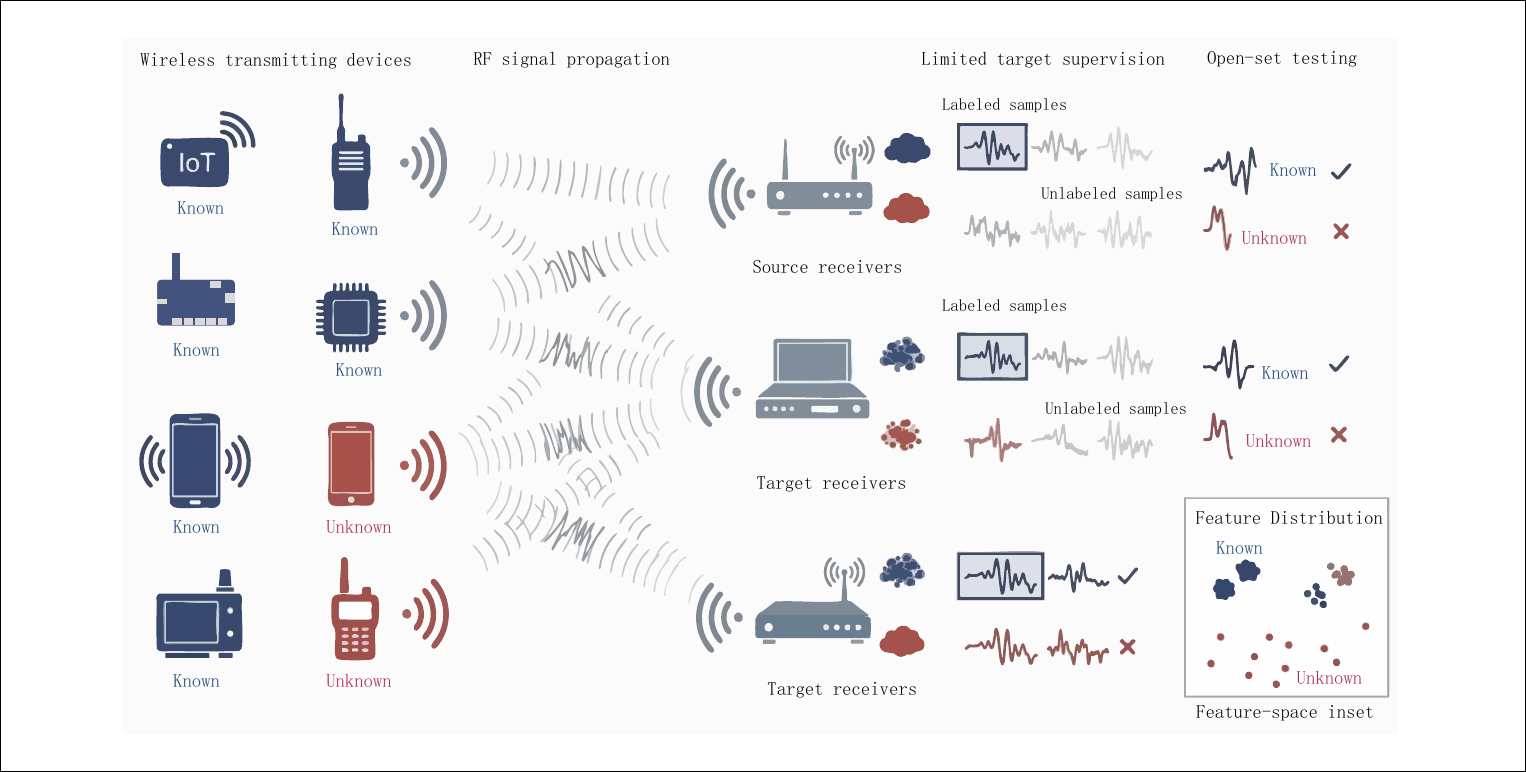}
\caption{Application scenario of cross-receiver open-set radio frequency fingerprint identification.}
\label{fig:intro-scenario}
\end{figure*}

\IEEEPARstart{R}{adio} frequency fingerprint identification (RFFI) provides a critical physical-layer credential for wireless ad hoc and Internet of Things (IoT) networks by authenticating devices through hardware-induced transmitter impairments~\cite{zhang2023rffi_iot_auth}.
Unlike centralized networks with uniform infrastructure, ad hoc environments are highly decentralized and dynamic.
As shown in Fig.~\ref{fig:intro-scenario}, an operational authentication system routinely encounters both registered and unregistered devices while capturing signals across heterogeneous receivers.
This deployment pattern imposes two strict criteria: the credential must remain reliable across physically dispersed, heterogeneous receivers, and it must decisively reject traffic from unregistered rogue nodes.
These receivers introduce localized distortions.
Such distortions arise from variations in local oscillators, sampling clocks, and RF front-end filters, and they continuously shift the in-phase/quadrature (I/Q) feature manifold~\cite{hanna2022wisig,shen2024receiver_agnostic,li2025zero_trust_receiver_agnostic,ma2025_receiver_independent}.
However, existing methods typically evaluate RFFI under an idealized closed-system assumption, whereas real-world ad hoc deployments suffer from severe condition mismatches.
Consequently, RFFI is fundamentally a cross-receiver transfer and unknown-device rejection problem, not merely a static closed-set classification task.

Prior work has advanced both domain adaptation and open-set recognition, but the two research lines remain only partially connected.
Cross-receiver studies characterize strong receiver shifts but often retain a strictly closed-set label space~\cite{hanna2022wisig,ma2025_receiver_independent}; under this setting, unknown transmitters may be classified as registered identities.
Conversely, many open-set methods evaluate rejection under the same receiver or weak domain shifts~\cite{huang2025medae}, although emerging cross-domain open-set RFFI has begun to bridge the two settings~\cite{hong2026pcpd}.
Practical IoT authentication nevertheless requires both capabilities, because a deployed receiver simultaneously observes shifted known features and entirely novel unknown identities.

To bridge this gap, this paper investigates the single-source single-target cross-receiver open-set setting.
This strict deployment protocol excludes unknown target transmitters from training, threshold calibration, and model selection.
As a result, the rejection decision must rely exclusively on source validation scores.
This rigid calibration exposes two contrasting failure modes in deep neural networks.
First, receiver shift depresses the confidence scores of target-known samples, causing \textbf{false rejection} of registered devices.
Second, naive closed-set domain alignment may draw unseen target transmitters toward known regions, increasing the risk of \textbf{false acceptance} for unregistered traffic.

Therefore, we approach this task not as standard domain adaptation, but as a threshold reliability problem.
To this end, we propose CRODA-ST, which jointly constrains structural semantics and decision boundaries through two complementary objectives rather than optimizing solely for classification accuracy.
Within this objective, Discriminative Structure Anchoring (DSA) reconstructs target-known reference points to establish a stable semantic foundation, while Rejection-Oriented Alignment (ROA) concurrently regularizes receiver-sensitive confidence boundaries.
By anchoring target-known geometry and regularizing confidence behavior in the same optimization, CRODA-ST improves the reliability of maximum softmax probability (MSP) scoring under severe receiver shift.
This work makes three main contributions.

\textit{1) Problem formulation:} We define the single-source single-target cross-receiver open-set RFFI problem under strict source-only threshold calibration.
We demonstrate that shifted known samples and unseen transmitters induce opposing threshold errors, explicitly separating threshold reliability from standard closed-set calibration.

\textit{2) Method design and methodological insight:} We propose CRODA-ST, a cross-receiver open-set adaptation framework that couples target-known structural anchoring with receiver-oriented confidence regularization within a joint training objective.
DSA geometrically anchors the target-known spatial references to establish a stable semantic foundation, while ROA concurrently regularizes receiver-sensitive confidence behavior to improve open-set separability.
The methodological insight is to impose these constraints jointly rather than through sequential processing.
Under the strict calibration setting above, this joint constraint provides a structural formulation for balancing domain alignment and open-set rejection in cross-receiver RFFI.

\textit{3) Experimental verification:} We systematically validate the framework on the WiSig ManyTx dataset, where CRODA-ST achieves a 0.9580 open-set classification rate (OSCR) while reducing false acceptance.
Beyond these WiFi empirical measurements, we evaluate the proposed architecture on a synthesized Chirp Spread Spectrum (CSS) LoRa dataset.
This simulation-based cross-protocol diagnostic tests whether the same open-set formulation remains usable under systematically controlled transmitter and receiver hardware impairments.

The remainder of this paper is organized as follows. Section~2 reviews RFFI-based IoT authentication, cross-receiver transfer, and open-set recognition, positioning the proposed setting relative to existing paradigms.
Section~3 formulates the cross-receiver open-set problem and details the DSA and ROA mechanisms of CRODA-ST.
Section~4 reports the experimental protocol, comparative analysis, ablations, and specialized evaluations on the LoRa array. Section~5 concludes the paper.

\section{Related Work}

\subsection{Radio Frequency Fingerprint Identification}
Radio Frequency Fingerprint Identification (RFFI) provides a physical-layer authentication mechanism by exploiting hardware-intrinsic imperfections in the radio-frequency front end of transmitters \cite{zhang2023rffi_iot_auth,sa2019sei_iot,yan2025rffi_review}.
As wireless devices are increasingly deployed in open IoT environments, RFFI has become an essential supplement to cryptographic protocols by offering spoofing-resistant and lightweight device verification \cite{peng2019hybrid_rffi_iot,cai2024lightweight_uav,qi2024lightweight_v2x}.
Traditional feature engineering and deep learning models have achieved promising classification accuracy under controlled same-domain scenarios, where the propagation channel and receiver hardware remain stable \cite{wang2025model_based_rff,shen2022_lora_rffi,xing2023channel_robust_rffi,peng2024_lte_channel_robust}.
Practical deployments, however, often involve diverse receiver hardware, changing acquisition environments, and unpredictable environmental noise \cite{hanna2022wisig,chillet2024_rifyfi}.
Such hardware and environmental variations induce significant distribution shifts, so representations learned under ideal assumptions transfer poorly and hinder the practical deployment of RFFI.

\subsection{Cross-Receiver and Domain Adaptation in RFFI}
To mitigate the deployment mismatch caused by hardware variations, cross-receiver RFFI has been formulated as a domain shift problem.
Prior studies address receiver-induced differences through domain adaptation, feature transformation, or domain generalization.
For example, adversarial training and contrastive alignment have been used to extract receiver-invariant representations \cite{shen2024receiver_agnostic,ma2025_receiver_independent,zhang2025_domain_generalization_crossrx,zha2023_subdomain_contrastive_crossrx}.
Other strategies incorporate feature transformation, pseudo-label self-training, and multi-objective domain adaptation to stabilize target-domain alignment \cite{yang2024_receiver_impact_da,xiao2025ftan,yang2025cscnet,feng2025dynamic_distribution}.
Although these cross-receiver methods mitigate hardware-induced domain shifts, most retain a closed-set assumption.
They generally assume that the source and target domains share identical label spaces.
In open IoT environments, however, target data may also contain unregistered or rogue devices.
Under this label-space mismatch, forcing distribution alignment may draw unknown-device features toward known-class regions.
This behavior can increase false acceptance and weaken the security role of RFFI.

\subsection{Open-Set RFFI}
Recognizing the limitations of closed-set classifiers, open-set recognition (OSR) requires the model to correctly classify known devices while rejecting unknown classes absent during training \cite{bendale2016openmax,geng2021osr_survey,chen2021_arpl}.
In the context of RFFI, recent methods improve unknown-class rejection using metric-enhanced autoencoders, prototype learning, out-of-distribution exposure, and test-time calibration \cite{huang2025medae,wang2025os_sei_oels,xing2025rff_osr_iot,huang2024_open_set_noisy,ma2025_multitask_prototype_openrfi,liu2026adaptive_weibull}.
While these open-set methods improve rejection under stable conditions, most of them are evaluated under the same receiver or weak domain shifts.
In cross-receiver deployments, hardware-induced feature drift distorts both the known-class boundaries and the confidence distributions used for rejection.
Existing rejection mechanisms, including distance thresholds and prototype matching, may consequently become unreliable when directly transferred to a new receiver.

\subsection{Open-Set Recognition Across Receivers}
In the broader machine learning community, unknown-aware alignment has been studied through open-set domain adaptation (OSDA) and universal domain adaptation, which explicitly consider label-space discrepancies across source and target domains \cite{gao2020_adv_osda,loghmani2020_pu_osda,wang2021_category_attention_osda,zhao2023_deconfounding_osda}.
Open Set Domain Adaptation by Backpropagation (OSBP) uses adversarial learning to separate target samples that do not belong to the source label set~\cite{saito2018_osbp}; Separate to Adapt (STA) progressively separates transferable and unknown target samples~\cite{liu2019_sta}; and universal domain adaptation extends this problem to more general label-space relationships~\cite{you2019_universal_da}.
These methods provide important precedents for unknown-aware adaptation, but their assumptions do not directly match cross-receiver RFFI deployment, and they have not been systematically tailored to suppress receiver-specific hardware artifacts in RFFI.
They typically exploit unlabeled target-domain samples during adaptation and infer target-private samples from prediction uncertainty or transferability scores.
The source of the shift is also different: receiver-induced distortions alter the I/Q feature geometry through oscillator offsets, front-end filtering, and related hardware effects, rather than through the image-domain appearance shifts usually studied in generic domain adaptation.
Under this receiver-induced domain shift, indiscriminate adversarial alignment may draw unknown transmitters toward known-device regions.

A small number of studies have begun to jointly model cross-domain transfer and open-set rejection.
For instance, Hong et al. explored cross-domain open-set RFFI through pseudo-label learning and prototype calibration \cite{hong2026pcpd}.
Generative auxiliary modeling and multi-task discrimination have also been introduced to combine feature alignment with open-set robustness \cite{guo2024acgan_openmax,yang2025_openrfi}.
Different from approaches that require abundant target-domain data or complex generative models, this work focuses on a highly constrained yet realistic scenario: cross-receiver open-set recognition with only limited labeled target-domain data and source-only threshold calibration.
This gap motivates a structural approach rather than a direct reuse of generic OSDA objectives.

\section{Proposed Method}

\subsection{Problem Formulation and Design Rationale}
We formulate single-source single-target cross-receiver open-set RFFI as a source-calibrated rejection problem.
The source receiver provides labeled known-class samples $\mathcal{D}_s=\{(x_i^s,y_i^s)\}_{i=1}^{n_s}$, where $y_i^s\in\mathcal{Y}_K=\{1,\ldots,K\}$ denotes a registered transmitter.
At the target receiver, adaptation is limited to the labeled enrollment set $\mathcal{D}_t^l=\{(x_j^t,y_j^t)\}_{j=1}^{n_t}$ from the same known transmitter set.
This limited set has a direct deployment meaning: it represents a low-cost onboarding step in IoT deployments.
During physical registration, each device contributes only a few milliseconds of I/Q signal capture.
Unknown target identities $\mathcal{Y}_U$ satisfy $\mathcal{Y}_K\cap\mathcal{Y}_U=\emptyset$ and appear only during testing.
Consequently, no target-unknown sample is used for training, threshold calibration, hyperparameter tuning, or model selection.

This protocol differs from ordinary closed-set adaptation because the final target decision depends on a Maximum Softmax Probability (MSP) threshold calibrated only from source known samples.
Such source-only calibration requires shifted target-known samples to remain compatible with the known-class score structure learned from the source receiver.
Otherwise, the threshold may reject registered transmitters for receiver-related rather than identity-related reasons.
At the same time, adaptation on enrolled target-known samples does not by itself guarantee rejection of unseen transmitters, which may still receive high known-class confidence.
CRODA-ST maps these two requirements onto the same encoder--classifier path: DSA anchors target-known structure for compatibility with source-calibrated rejection, whereas ROA addresses receiver-identifiable variation in known-sample embeddings and local posterior sensitivity that can destabilize rejection scores.

\begin{itemize}
    \item \textbf{Discriminative Structure Anchoring (DSA) against Feature Drift:} Rebuilds target-side references for registered transmitters from labeled target enrollment samples.
    These references anchor where known classes should lie under the target receiver, with the aim of counteracting receiver-induced feature drift and limiting receiver-induced false rejection.
    \item \textbf{Rejection-Oriented Alignment (ROA) against Confidence Inflation:} Concurrently regularizes receiver-sensitive variations in known-sample embeddings and posteriors strictly around the anchored known-class structure.
    This confidence regularization is intended to reduce confidence inflation, receiver-induced score fluctuations, and the associated false-acceptance risk without altering the inference rule.
\end{itemize}

Without concurrent anchoring, confidence regularization alone leaves the target-known structure insufficiently constrained under receiver shift.
Target-unknown transmitters may therefore still receive high confidence in known regions during testing.
Conversely, anchoring the known-class features alone can recover target-domain recognition accuracy, while rejection scores remain sensitive to receiver-induced confidence fluctuations.
CRODA-ST therefore couples DSA and ROA within a single structure-anchored objective: DSA constructs a target-anchored known-class foundation, whereas ROA regularizes confidence behavior around this foundation.
This complementarity between target-known structural anchoring and rejection-oriented regularization is the central methodological contribution of CRODA-ST.

The corresponding losses are jointly optimized in every epoch on source samples and labeled target enrollment samples.
The complete training, calibration, and inference procedure is summarized in Algorithm~\ref{alg:croda-st}.

\subsection{Input Representation and Encoder}
The input is a complex baseband segment represented by in-phase and quadrature (I/Q) channels.
Energy normalization mitigates sample-scale variation, while source-fitted standardization applies the same reference statistics to source training, target enrollment, source validation, and test samples.
We write the resulting I/Q--time representation as $x\in\mathbb{R}^{2\times256}$.

A residual backbone $b_{\theta}(\cdot)$ maps $x$ to a feature that should preserve transmitter-discriminative structure while reducing sensitivity to receiver-dependent distortion.
A linear projection with learnable parameters $W_p$ and $b_p$ completes the encoder $f_{\theta}(\cdot)$ and produces the shared 128-D embedding $z$ used by all downstream losses and by the common classifier/MSP recognition and rejection path:
\begin{equation}
z=f_{\theta}(x)=W_p b_{\theta}(x)+b_p,\qquad z\in\mathbb{R}^{128}.
\end{equation}

A single linear $K$-way classifier with parameters $W_c$ and $b_c$ maps the embedding to known-class logits and posteriors:
\begin{equation}
\ell(x)=W_c z+b_c,\qquad p(x)=\mathrm{softmax}(\ell(x)).
\end{equation}
Here, $p_k(x)$ denotes the posterior probability assigned to known class $k$.

Thus, recognition and open-set scoring use the same 128-D embedding and the same linear classifier.
CRODA-ST keeps the inference rule MSP-based. Its rejection gain therefore comes from target-known anchoring and confidence regularization, rather than from post-hoc OpenMax recalibration or dummy unknown classes.

\subsection{Discriminative Structure Anchoring (DSA)}
Source-calibrated rejection is meaningful only if the classifier's known-class decision surface remains reliable for both receivers.
Blind feature adaptation may let target representations drift. DSA therefore treats the scarce target enrollment samples as spatial ``anchors'' rather than weak auxiliary data.
It enforces two constraints. First, the decision surface must classify both source and target known samples correctly.
Second, source and target embeddings of the \textit{same} transmitter must form a shared local cluster, yielding a compact known-class space.
This gives target-known samples the class structure required by MSP-based rejection.

The source-side component applies a standard cross-entropy loss over source mini-batches $\mathcal{B}_s\subset\mathcal{D}_s$ to maintain the discriminative surface:
\begin{equation}
\mathcal{L}_{s}=-\frac{1}{|\mathcal{B}_s|}\sum_{(x_i^s,y_i^s)\in\mathcal{B}_s}\log p_{y_i^s}(x_i^s).
\end{equation}

Its paired target-side component applies the same loss over target enrollment mini-batches $\mathcal{B}_t^l\subset\mathcal{D}_t^l$:
\begin{equation}
\mathcal{L}_{t}=-\frac{1}{|\mathcal{B}_t^l|}\sum_{(x_j^t,y_j^t)\in\mathcal{B}_t^l}\log p_{y_j^t}(x_j^t).
\end{equation}

Together, $\mathcal{L}_{s}$ and $\mathcal{L}_{t}$ form the supervised DSA pair, complemented by an embedding-level clustering constraint.
We merge the samples into $\mathcal{B}_{st}=\mathcal{B}_s\cup\mathcal{B}_t^l$ and normalize each embedding as $\tilde{z}=z/\|z\|_2$.
For anchor $i$, $\mathcal{P}(i)=\{j:j\ne i,\,y_j=y_i\}$ denotes its positive set, and $\mathcal{A}=\{i:|\mathcal{P}(i)|>0\}$ collects anchors with at least one positive.
The supervised contrastive loss pulls source and target embeddings of the same transmitter together while separating different transmitters:
\begin{equation}
\mathcal{L}_{\mathrm{sup}}=-\frac{1}{|\mathcal{A}|}\sum_{i\in\mathcal{A}}\frac{1}{|\mathcal{P}(i)|}\sum_{j\in\mathcal{P}(i)}\log\frac{\exp(\tilde{z}_i^{\top}\tilde{z}_j/T_c)}{\sum_{a\in\mathcal{B}_{st}\setminus\{i\}}\exp(\tilde{z}_i^{\top}\tilde{z}_a/T_c)},
\end{equation}
where $T_c$ is a scalar temperature parameter that controls the concentration of the distribution.
Together with the supervised DSA pair, this term supports the compact known-class space required by MSP-based rejection.

\subsection{Rejection-Oriented Alignment (ROA)}
DSA anchors the target known-class references, yet receiver-dependent hardware variations may still perturb posterior distributions and MSP scores near rejection boundaries.
ROA addresses this remaining confidence risk by stabilizing confidence scores around the anchored structure while leaving the inference path unchanged.
Its Gradient Reversal Layer (GRL)-based receiver discriminator~\cite{ganin2016_domain_adversarial} is trained to reduce receiver-identifiable information in known-sample embeddings; local oscillator (LO) offsets and radio-frequency (RF) front-end filter distortions are examples of receiver hardware differences that may contribute such information. Its effect on false acceptance is evaluated in the ablation study.
In parallel, Virtual Adversarial Training (VAT)~\cite{miyato2018_vat} encourages local posterior smoothness under small adversarial input perturbations and regularizes the sensitivity of MSP scores used for rejection.

For each known sample $x\in\mathcal{B}_s\cup\mathcal{B}_t^l$, ROA passes the normalized embedding $\tilde{z}$ through the GRL to the receiver discriminator $d_{\phi}(\cdot)$.
Let $q(x)=\mathrm{softmax}(d_{\phi}(\mathrm{GRL}(\tilde{z})))$ and let $r(x)\in\{0,1\}$ denote the receiver label. The source--target receiver loss is calculated exclusively on known samples:
\begin{equation}
\mathcal{L}_{\mathrm{dom}}=-\frac{1}{|\mathcal{B}_s|+|\mathcal{B}_t^l|}\sum_{x\in\mathcal{B}_s\cup\mathcal{B}_t^l}\log q_{r(x)}(x).
\end{equation}
During backpropagation, the GRL reverses encoder-side gradients, encouraging embeddings that are less predictive of receiver identity.

Local confidence stability requires the posterior to remain smooth under small signal perturbations.
In RFFI, local input sensitivity is relevant because hardware noise may perturb signal phase and amplitude; however, the optimization-derived $r_{\mathrm{adv}}$ represents a worst-case local direction rather than a calibrated physical-noise model.
The virtual adversarial direction is $r_{\mathrm{adv}}(x)=\arg\max_{\lVert r\rVert_2\le\epsilon}D_{\mathrm{KL}}(p(x)\|p(x+r))$, where $\epsilon$ is the $L_2$ perturbation radius and $D_{\mathrm{KL}}$ denotes the Kullback--Leibler (KL) divergence. It is approximated using one power iteration initialized at $\xi=10^{-6}$.
VAT then compares the fixed reference posterior $p(x)$ with the perturbed posterior $p(x+r_{\mathrm{adv}})$ through the following smoothing loss:
\begin{equation}
\mathcal{L}_{\mathrm{vat}}=\frac{1}{|\mathcal{B}_{st}|}\sum_{x\in\mathcal{B}_{st}}D_{\mathrm{KL}}\!\left(p(x)\,\middle\|\,p(x+r_{\mathrm{adv}})\right).
\end{equation}
This term encourages local posterior consistency under adversarial input perturbations.

\subsection{Optimization and Deployment Output}
The complete objective combines the DSA terms $\mathcal{L}_{s}$, $\mathcal{L}_{t}$, and $\mathcal{L}_{\mathrm{sup}}$, which anchor registered-transmitter structure across the two receivers, with the ROA terms $\mathcal{L}_{\mathrm{dom}}$ and $\mathcal{L}_{\mathrm{vat}}$, which regularize receiver-sensitive confidence changes around that structure:
\begin{equation}
\mathcal{L}=\mathcal{L}_{s}+\lambda_t\mathcal{L}_{t}+\lambda_c\mathcal{L}_{\mathrm{sup}}+\lambda_d\mathcal{L}_{\mathrm{dom}}+\lambda_v\mathcal{L}_{\mathrm{vat}}.
\end{equation}
The nonnegative coefficients $\lambda_t$, $\lambda_c$, $\lambda_d$, and $\lambda_v$ weight target supervision, supervised contrast, receiver alignment, and local smoothing, respectively.
All terms are optimized jointly in every training epoch.

At deployment, only the trained encoder $f_{\theta}(\cdot)$ and the linear $K$-way classifier are retained.
For an input $x$, the classifier posterior $p(x)$ defines the knownness score and raw known-class prediction:
\begin{equation}
s(x)=\max_{k\in\mathcal{Y}_K}p_k(x),\qquad\hat{k}(x)=\arg\max_{k\in\mathcal{Y}_K}p_k(x).
\end{equation}

Crucially, the threshold is estimated \textit{only} from the source known-class validation set $\mathcal{D}_{\mathrm{val}}^s$:
\begin{equation}
\tau=Q_{1-\rho}\left(\{s(x_i):(x_i,y_i)\in\mathcal{D}_{\mathrm{val}}^s\}\right).
\end{equation}
Here, $Q_{\alpha}(\cdot)$ denotes the empirical $\alpha$-quantile. The retention rate $\rho$ sets a source-side operating point: the threshold accepts approximately a fraction $\rho$ of registered source-validation samples.
The final target rule then rejects unknown transmitters at deployment:
\begin{equation}
\hat{y}(x)=
\begin{cases}
\hat{k}(x), & s(x)\ge \tau,\\
-1, & s(x)<\tau.
\end{cases}
\end{equation}
Here, the output label $-1$ denotes rejection as an unknown transmitter.

\begin{algorithm}[!t]
\caption{Training and Deployment Procedure of CRODA-ST}
\label{alg:croda-st}
\begin{algorithmic}[1]
\REQUIRE Source training set $\mathcal{D}_s$, Target enrollment set $\mathcal{D}_t^l$, Source validation set $\mathcal{D}_{\mathrm{val}}^s$.
\REQUIRE Target test sample $x_t$, Hyperparameters $\lambda_t, \lambda_c, \lambda_d, \lambda_v, T_c$, Retention rate $\rho$.
\STATE \textbf{\# Phase 1: Joint Structural Anchoring and Alignment}
\FOR{each epoch}
    \STATE Sample mini-batches $\mathcal{B}_s \sim \mathcal{D}_s$ and $\mathcal{B}_t^l \sim \mathcal{D}_t^l$.
    \STATE Compute cross-entropy losses $\mathcal{L}_s$ and $\mathcal{L}_t$ via Eq.~(3) and (4).
    \STATE Compute supervised contrastive loss $\mathcal{L}_{\mathrm{sup}}$ with temperature $T_c$ via Eq.~(5).
    \STATE Compute adversarial alignment loss $\mathcal{L}_{\mathrm{dom}}$ using GRL via Eq.~(6).
    \STATE Estimate perturbation $r_{\mathrm{adv}}$ and compute VAT smoothing loss $\mathcal{L}_{\mathrm{vat}}$ via Eq.~(7).
    \STATE Update network parameters to minimize total objective $\mathcal{L}$ in Eq.~(8).
\ENDFOR
\STATE \textbf{\# Phase 2: Source-Only Unknown-Free Calibration}
\FOR{each $(x_i, y_i) \in \mathcal{D}_{\mathrm{val}}^s$}
    \STATE Extract prediction posterior $p(x_i)$ and compute knownness score $s(x_i) = \max_{k} p_k(x_i)$.
\ENDFOR
\STATE Calculate threshold $\tau$ corresponding to the retention rate $\rho$ via Eq.~(10).
\STATE \textbf{\# Phase 3: Online Deployment Inference}
\STATE Forward test sample $x_t$ through the encoder and classifier.
\STATE Compute test knownness score $s(x_t) = \max_{k} p_k(x_t)$.
\IF{$s(x_t) \ge \tau$}
    \STATE \textbf{Output} Known identity $\hat{y}(x_t) = \arg\max_{k} p_k(x_t)$.
\ELSE
    \STATE \textbf{Output} Rejected as target-unknown identity ($\hat{y}(x_t) = -1$).
\ENDIF
\end{algorithmic}
\end{algorithm}

\section{Experimental Results and Analysis}

\subsection{Experimental Setup}
We systematically evaluate the proposed CRODA-ST framework on the large-scale WiSig ManyTx WiFi baseband dataset~\cite{hanna2022wisig}.
This testbed provides heterogeneous receivers, many transmitters, and cross-day captures. These conditions mirror real-world IoT deployment mismatches.
The main text reports the canonical WiSig transfer setting and temporal enrollment tests under the same calibration protocol.
The canonical setting uses receiver 1-1 as the source receiver and receiver 1-19 as the target receiver, with 37 known training classes among 114 target-test classes (openness 0.30), 20 labeled target-enrollment samples per known class, three class splits, and three random seeds.
Following the standard open-set definition, openness is computed as $\mathcal{O}=1-\sqrt{2K_{\rm tr}/(K_{\rm tr}+K_{\rm te})}$, where $K_{\rm tr}$ is the number of known training classes and $K_{\rm te}$ is the total number of test classes.
To complement these empirical WiFi evaluations, Appendix~\ref{app:supplementary-experiments} reports receiver-pair robustness over 10 source-target pairs, openness sensitivity, online inference cost, retention sensitivity, and synthetic LoRa analyses.

For data representation, each length-256 I/Q sample is centered, energy-normalized, standardized using source-training statistics, and formatted as a $2\times256$ input tensor.
The backbone architecture employs a 2-D residual I/Q--time encoder with 32 base channels, $(2,2,2)$ residual blocks, and a 128-dimensional classification embedding.
During optimization, we train the network using the Adam optimizer for 80 epochs.
The batch size is 64, the learning rate is $10^{-3}$, and the weight decay is $10^{-4}$, with cosine annealing.
Unless otherwise specified, the hyperparameters are $\lambda_t=1.0$, $\lambda_{\mathrm{c}}=0.05$, $T_{\mathrm{c}}=0.07$, $\lambda_{\mathrm{d}}=0.1$, $\lambda_{\mathrm{v}}=0.05$, the VAT $L_2$ perturbation radius $\epsilon=2.0$, and a gradient reversal coefficient of $1.0$.
For receiver-pair robustness, each source-target pair in Appendix~\ref{app:receiver-pair-robustness} is averaged over the same three class splits and three random seeds; these pair-level results are reported separately rather than pooled into the main comparison.

The reported evaluation metrics comprehensively capture the closed-set recognition and open-set rejection tradeoff: known-class accuracy ($\mathrm{Acc}_{K}$), area under the receiver operating characteristic curve (AUROC), open-set classification rate (OSCR), and the false positive rate at a 90\% true positive rate ($\mathrm{FPR}_{90}$).
Let $\mathcal{D}_{K}$ and $\mathcal{D}_{U}$ denote the known and unknown test sets, respectively.
For a known test sample $x$ and its ground-truth label $y$, let $s(x)$ denote the knownness score, $\hat{k}(x)$ the predicted known class, and $\mathbb{I}[\cdot]$ the indicator function.
Throughout the receiver operating characteristic analysis, known samples are treated as positives and unknown samples as negatives; $\gamma$ denotes a generic score threshold.
The fundamental probabilities---true positive rate ($\mathrm{TPR}$), false positive rate ($\mathrm{FPR}$), and correct classification rate ($\mathrm{CCR}$)---are formulated as follows:
\begin{equation}
\scalebox{0.88}{$
\begin{aligned}
\mathrm{Acc}_{K}
&=\frac{1}{|\mathcal{D}_{K}|}\sum_{(x,y)\in\mathcal{D}_{K}}
\mathbb{I}[\hat{k}(x)=y],\\
\mathrm{TPR}(\gamma)
&=\frac{1}{|\mathcal{D}_{K}|}\sum_{x\in\mathcal{D}_{K}}
\mathbb{I}[s(x)\ge\gamma],\\
\mathrm{FPR}(\gamma)
&=\frac{1}{|\mathcal{D}_{U}|}\sum_{x\in\mathcal{D}_{U}}
\mathbb{I}[s(x)\ge\gamma],\\
\mathrm{CCR}(\gamma)
&=\frac{1}{|\mathcal{D}_{K}|}\sum_{(x,y)\in\mathcal{D}_{K}}
\mathbb{I}[\hat{k}(x)=y,\,s(x)\ge\gamma].
\end{aligned}
$}
\end{equation}
These metrics capture complementary aspects of cross-receiver open-set authentication. $\mathrm{Acc}_{K}$ measures closed-set identity recognition on registered transmitters;
a higher value means that target-known samples are assigned to the correct known identities before rejection is considered.
$\mathrm{TPR}(\gamma)$ measures the fraction of known samples accepted as known at score threshold $\gamma$, whereas $\mathrm{FPR}(\gamma)$ measures the fraction of unknown samples incorrectly accepted as known.
$\mathrm{CCR}(\gamma)$ is stricter than $\mathrm{TPR}(\gamma)$ because a known sample contributes only when it is both accepted and correctly classified.

AUROC summarizes score separation by integrating the $(\mathrm{FPR}(\gamma),\mathrm{TPR}(\gamma))$ curve. OSCR integrates the $(\mathrm{FPR}(\gamma),\mathrm{CCR}(\gamma))$ curve and therefore rewards methods that preserve known-class discrimination while rejecting unknown traffic.
Together, AUROC and OSCR separate score-level rejection from class-correct rejection.
The metric $\mathrm{FPR}_{90}$ is defined as $\mathrm{FPR}(\gamma_{90})$, where $\gamma_{90}$ is selected on the known test scores within the evaluation domain such that $\mathrm{TPR}(\gamma_{90})=0.90$.
Accordingly, $\mathrm{FPR}_{90}$ measures the fraction of unknown samples in that same domain accepted at a threshold that retains 90\% of its known samples.
The deployed threshold $\tau$ is instead calibrated on source validation known samples according to Eq.~(10). Its false acceptance rate (FAR) is $\mathrm{FAR}(\tau)=|\mathcal{D}_{U}|^{-1}\sum_{x\in\mathcal{D}_{U}}\mathbb{I}[s(x)\ge\tau]=\mathrm{FPR}(\tau)$. We reserve $\mathrm{FPR}_{90}$ for $\gamma_{90}$ and use deployment FAR only for the source-calibrated threshold $\tau$.

\subsection{Compared Baselines and Protocol Adaptation}
Because dedicated cross-receiver open-set RFFI frameworks remain scarce in the literature, we construct a comprehensive comparison using representative methods adapted from related fields.
Specifically, we compare CRODA-ST with post-hoc calibration, metric-based open-set modeling, and domain-adaptation baselines under the same deployment-constrained protocol.
The first category includes methods that do not explicitly perform cross-receiver feature alignment.
OpenMax w/ target enrollment~\cite{bendale2016openmax} calibrates the network by fitting class-specific Weibull tail models to known-class activation distances and reallocating activation mass to an explicit unknown class.
MeDAE~\cite{huang2025medae} represents metric-based open-set modeling, fitting a metric-enhanced denoising autoencoder with the allowed target-enrollment samples to compress intra-class features and reject samples far from established known-class centers.

The second category includes methods that explicitly address distribution shift, mainly through objectives originally designed for unsupervised domain adaptation (UDA).
PCPD~\cite{hong2026pcpd} is a prototype-calibrated open-set adaptation framework that employs adversarial domain alignment and high-confidence target pseudo-labeling.
The Feature Transform and Alignment Network (FTAN)~\cite{xiao2025ftan} serves as a representative closed-set UDA baseline, aligning source and target representations through cross-domain feature conversion, maximum mean discrepancy (MMD), and target self-training.
Since FTAN does not originally include an unknown-rejection branch, we apply the same MSP rule for open-set scoring.
Finally, the Domain-Adversarial Neural Network with Weibull calibration (DANN+Weibull) adapts the cross-receiver open-set pipeline from~\cite{liu2025open_set_cross_receiver}, first learning receiver-invariant features using a GRL~\cite{ganin2016_domain_adversarial} and then fitting Weibull extreme value theory (EVT) models over source known-class feature distances for unknown rejection.
Although these UDA-style formulations do not originally rely on target-domain labels, we fit their adaptation or calibration branches with the same labeled target-enrollment set allowed by our protocol.

\begin{table}[!t]
\caption{Summary of Baseline Adaptation Protocols}
\label{tab:baseline-protocol}
\centering
\scriptsize
\setlength{\tabcolsep}{3pt}
\renewcommand{\arraystretch}{1.1}
\begin{tabular}{@{}lcccc@{}}
\toprule
Method & \makecell{Target\\Enrollment\\Labels Used?} & \makecell{Target\\Unknown\\in Training?} & \makecell{Rejection\\Score} & \makecell{Threshold\\Calibration} \\
\midrule
OpenMax & Yes (Fit) & No & Weibull & Source \\
MeDAE & Yes (Fit) & No & Distance & Source \\
PCPD & Yes (Fit) & No & Prototype & Source \\
FTAN & Yes (Fit) & No & MSP & Source \\
DANN+Weibull & Yes (Fit) & No & Weibull & Source \\
\textbf{CRODA-ST} & \textbf{Yes (Anchor)} & \textbf{No} & \textbf{MSP} & \textbf{Source} \\
\bottomrule
\end{tabular}
\end{table}

For fairness, all baselines share the same source split, target enrollment budget, input representation, and threshold-calibration rule as CRODA-ST.
We use the same base encoder wherever compatible and retain method-specific modules when the original objective requires them.
Table~\ref{tab:baseline-protocol} details which method-specific branches use the target-enrollment labels.
For UDA-style baselines whose original papers assume unlabeled target-domain data, this label use is a protocol adaptation to the target-enrollment setting rather than a change to the deployment calibration rule.
All deployment thresholds are calibrated from source validation known samples.

\subsection{Overall Cross-Receiver Comparison}
\begin{table*}[!t]
\caption{Target- and source-domain comparison in the canonical cross-receiver open-set setting.}
\label{tab:st-main-comparison-target}
\centering
\scriptsize
\setlength{\tabcolsep}{5pt}
\renewcommand{\arraystretch}{1.05}
\begin{tabular}{lcccc}
\toprule
Method & $\mathrm{Acc}_{K}$ $\uparrow$ & AUROC $\uparrow$ & OSCR $\uparrow$ & $\mathrm{FPR}_{90}$ $\downarrow$ \\
\midrule
\multicolumn{5}{l}{\textit{Target-domain test}} \\
OpenMax w/ target enrollment~\cite{bendale2016openmax} & 0.7664$\pm$0.0142 & 0.8328$\pm$0.0145 & 0.7434$\pm$0.0129 & 0.7579$\pm$0.0492 \\
MeDAE~\cite{huang2025medae} & 0.8449$\pm$0.0319 & \dashuline{0.9521$\pm$0.0159} & \dashuline{0.9313$\pm$0.0252} & \dashuline{0.1004$\pm$0.0657} \\
PCPD~\cite{hong2026pcpd} & 0.4365$\pm$0.1146 & 0.8557$\pm$0.0262 & 0.6668$\pm$0.2260 & 0.4638$\pm$0.0520 \\
FTAN~\cite{xiao2025ftan} & 0.8867$\pm$0.0330 & 0.9070$\pm$0.0186 & 0.8749$\pm$0.0264 & 0.2976$\pm$0.0767 \\
DANN+Weibull~\cite{liu2025open_set_cross_receiver} & \dashuline{0.9005$\pm$0.0210} & 0.8905$\pm$0.0329 & 0.8685$\pm$0.0364 & 0.2970$\pm$0.0849 \\
CRODA-ST & \textbf{0.9092$\pm$0.0182} & \textbf{0.9692$\pm$0.0099} & \textbf{0.9580$\pm$0.0122} & \textbf{0.0469$\pm$0.0309} \\
\midrule
\multicolumn{5}{l}{\textit{Source-domain test}} \\
OpenMax w/ target enrollment~\cite{bendale2016openmax} & 0.8851$\pm$0.0173 & 0.9129$\pm$0.0167 & 0.8661$\pm$0.0194 & 0.2572$\pm$0.1443 \\
MeDAE~\cite{huang2025medae} & \textbf{0.9451$\pm$0.0089} & \textbf{0.9808$\pm$0.0075} & \textbf{0.9726$\pm$0.0113} & \dashuline{0.0268$\pm$0.0225} \\
PCPD~\cite{hong2026pcpd} & 0.5662$\pm$0.3303 & 0.8183$\pm$0.1776 & 0.6659$\pm$0.3670 & 0.3900$\pm$0.3450 \\
FTAN~\cite{xiao2025ftan} & 0.9357$\pm$0.0067 & 0.9358$\pm$0.0136 & 0.9245$\pm$0.0157 & 0.1464$\pm$0.0518 \\
DANN+Weibull~\cite{liu2025open_set_cross_receiver} & 0.9297$\pm$0.0189 & 0.9194$\pm$0.0296 & 0.9025$\pm$0.0337 & 0.2120$\pm$0.0965 \\
CRODA-ST & \dashuline{0.9443$\pm$0.0092} & \dashuline{0.9794$\pm$0.0073} & \dashuline{0.9724$\pm$0.0090} & \textbf{0.0261$\pm$0.0141} \\
\bottomrule
\end{tabular}
\vspace{0.8mm}
\parbox{0.95\textwidth}{\scriptsize\emph{Note:} This table reports the canonical 1-1$\rightarrow$1-19 setting with 37 known training classes among 114 target-test classes, 20 labeled target-enrollment samples per known class, and 9 runs from three class splits and three random seeds. All methods follow the same source split, target-enrollment budget, and source-side threshold calibration.
OSCR: open-set classification rate. For $\mathrm{FPR}_{90}$, $\gamma_{90}$ is selected within each evaluation domain to retain 90\% of known test samples and is then applied to that domain's unknown test samples. It is distinct from the source-calibrated deployment threshold $\tau$.}
\end{table*}

In the evaluated target-domain setting, CRODA-ST leads all four metrics in Table~\ref{tab:st-main-comparison-target}.
Relative to the strongest baseline for each metric, it improves known-class accuracy by 0.0087 over DANN+Weibull, AUROC by 0.0171 and OSCR by 0.0267 over MeDAE, and reduces $\mathrm{FPR}_{90}$ by 0.0535 over MeDAE, corresponding to a 53.3\% relative reduction.
Within this setting, the metric-wise pattern supports a joint improvement in registered-device recognition and unknown rejection rather than a gain confined to either side of the tradeoff.

High target known-class accuracy alone, however, does not guarantee a low unknown acceptance rate.
DANN+Weibull and FTAN retain known-class accuracies of 0.9005 and 0.8867, respectively, while their $\mathrm{FPR}_{90}$ values remain at 0.2970 and 0.2976.
MeDAE lowers $\mathrm{FPR}_{90}$ to 0.1004, but its known-class accuracy decreases to 0.8449.
These results reveal a persistent recognition--rejection tradeoff for the compared adaptation and metric-learning baselines under receiver shift, without by themselves identifying a specific geometric failure mechanism.

OpenMax and PCPD exhibit larger target-domain recognition losses, reaching known-class accuracies of 0.7664 and 0.4365, respectively.
PCPD also shows substantial variability across runs ($\pm$0.1146).
Because PCPD relies on high-confidence target pseudo-labels, this instability is consistent with sensitivity to initial target-domain misalignment; nevertheless, the aggregate table alone does not isolate pseudo-label errors as its sole cause.

On the target-domain test, CRODA-ST combines a target known-class accuracy of 0.9092 with a target $\mathrm{FPR}_{90}$ of 0.0469.
On the source domain, MeDAE is marginally higher in known-class accuracy, AUROC, and OSCR, whereas CRODA-ST remains close on these metrics and achieves the lowest source-domain $\mathrm{FPR}_{90}$ of 0.0261.
At the source-calibrated deployment operating point $\rho=0.80$, CRODA-ST yields a target-unknown deployment FAR of 0.0075$\pm$0.0072 (Appendix~\ref{app:retention-sensitivity}).
The target- and source-domain $\mathrm{FPR}_{90}$ values are ranking-based within their respective evaluation domains, whereas deployment FAR uses the source-calibrated threshold $\tau$; these distinct operating conditions together support the scoped conclusion that the complete CRODA-ST configuration improves the recognition--rejection balance in the evaluated cross-receiver setting.

\subsection{Matched-Protocol Component Analysis and Mechanism Diagnostics}
\begin{table}[!t]
\caption{Matched-protocol target-domain component and full-model comparison in the canonical setting.}
\label{tab:st-mechanism-ablation}
\centering
\tiny
\renewcommand{\arraystretch}{0.76}
\resizebox{0.96\columnwidth}{!}{%
\begin{tabular}{lccc}
\toprule
Setting & $\mathrm{Acc}_{K}$ $\uparrow$ & OSCR $\uparrow$ & $\mathrm{FPR}_{90}$ $\downarrow$ \\
\midrule
\textsc{DSA w/o ROA} & 0.8512$\pm$0.0314 & 0.8996$\pm$0.0164 & 0.2360$\pm$0.0571 \\
\textsc{ROA w/o DSA} & 0.1483$\pm$0.0459 & 0.2300$\pm$0.0451 & 0.6021$\pm$0.1427 \\
\midrule
\textsc{Target Class Anchoring} & 0.8394$\pm$0.0337 & 0.8755$\pm$0.0235 & 0.2970$\pm$0.0725 \\
\textsc{Cross-Receiver Contrast} & 0.3520$\pm$0.1344 & 0.7956$\pm$0.0771 & 0.4524$\pm$0.1188 \\
\textsc{Full DSA} & 0.8512$\pm$0.0314 & 0.8996$\pm$0.0164 & 0.2360$\pm$0.0571 \\
\midrule
\textsc{DSA+Receiver Alignment} & 0.8474$\pm$0.0269 & 0.9002$\pm$0.0206 & 0.2275$\pm$0.0693 \\
\textsc{DSA+Local Smoothing} & 0.8629$\pm$0.0303 & 0.9156$\pm$0.0266 & 0.1627$\pm$0.0920 \\
\textsc{Full Objective} & \textbf{0.9092$\pm$0.0182} & \textbf{0.9580$\pm$0.0122} & \textbf{0.0469$\pm$0.0309} \\
\bottomrule
\end{tabular}%
}
\vspace{0.2mm}
\parbox{0.96\columnwidth}{\tiny\emph{Note:} All rows use the same canonical protocol and report means and standard deviations over the same 9 split--seed pairs (3 class splits $\times$ 3 seeds). The partial variants come from the loss-isolation sweep, whereas the full-objective row is evaluated from the same nine checkpoints used in the main comparison and therefore reflects the finalized loss weights. The progression is a matched-protocol comparison rather than a strict one-factor estimate. AUROC is omitted. \textsc{Full DSA} is functionally equivalent to \textsc{DSA w/o ROA}, combining Target Class Anchoring ($\mathcal{L}_t$) and Cross-Receiver Contrast ($\mathcal{L}_{\mathrm{sup}}$).
OSCR: open-set classification rate; $\mathrm{FPR}_{90}$: false positive rate at 90\% true positive rate, distinct from the deployment threshold $\tau$.}
\vspace{2mm}
\end{table}

Component-level results in Table~\ref{tab:st-mechanism-ablation} compare the partial objectives with the final full model under the same canonical splits and seeds.
Because the partial variants use the loss-isolation sweep whereas the full model uses the finalized main-series loss weights, the progression is interpreted as configuration-level evidence rather than a strict one-factor effect.

The ROA-without-DSA configuration records a known-class accuracy of 0.1483$\pm$0.0459. DSA alone records 0.8512, while its $\mathrm{FPR}_{90}$ remains 0.2360.
Within the DSA breakdown, Target Class Anchoring records a known-class accuracy of 0.8394 and an OSCR of 0.8755, whereas Cross-Receiver Contrast records 0.3520 and 0.7956, respectively. By design, target class anchoring supplies explicit spatial references. The contrast-only result shows that Cross-Receiver Contrast is insufficient for target recognition in isolation; it is not an independent contribution estimate.

Full DSA records an OSCR of 0.8996, compared with 0.9002 for DSA+Receiver Alignment. Among the two DSA-plus-single-ROA configurations, DSA+Local Smoothing records the larger numerical difference from Full DSA, with an OSCR of 0.9156 and an $\mathrm{FPR}_{90}$ of 0.1627. The finalized full model records a known accuracy of 0.9092, an OSCR of 0.9580, and an $\mathrm{FPR}_{90}$ of 0.0469. Relative to DSA+Local Smoothing, the numerical differences are +0.0463 in known accuracy, +0.0424 in OSCR, and $-$0.1157 in $\mathrm{FPR}_{90}$.

Taken together, these diagnostic comparisons are consistent with complementary functional contributions under the tested protocol. Because the finalized full model uses the main-series loss weights while the partial variants come from the loss-isolation sweep, the progression supports the complete configuration but remains a configuration-level comparison rather than a strict one-factor causal estimate.

To examine the proposed mechanism beyond aggregate ranking metrics, we evaluate the DSA-only checkpoints and the final main-series full-model checkpoints under the source-calibrated threshold at $\rho=0.80$ and measure their original 128-dimensional embeddings. Table~\ref{tab:st-mechanism-diagnostics} reports paired diagnostics over the same 9 split--seed combinations.

\begin{table*}[!t]
\caption{Source-calibrated and feature-geometry diagnostics for paired DSA-only and final full-model checkpoints.}
\label{tab:st-mechanism-diagnostics}
\centering
\scriptsize
\setlength{\tabcolsep}{5pt}
\renewcommand{\arraystretch}{0.95}
\resizebox{0.92\textwidth}{!}{%
\begin{tabular}{lccccc}
\toprule
Variant & \makecell{Known\\Reject@${\tau}$ $\downarrow$} & \makecell{Unknown\\FAR@${\tau}$ $\downarrow$} & \makecell{MSP\\$W_1$ $\downarrow$} & \makecell{Within-Class\\Cos. Dist. $\downarrow$} & \makecell{Unknown\\Intrusion $\downarrow$} \\
\midrule
DSA only & 0.3771$\pm$0.1015 & 0.0143$\pm$0.0072 & 0.0253$\pm$0.0080 & 0.0230$\pm$0.0052 & 0.6859$\pm$0.1357 \\
Full objective & \textbf{0.2498$\pm$0.0434} & \textbf{0.0075$\pm$0.0072} & \textbf{0.0112$\pm$0.0052} & \textbf{0.0146$\pm$0.0030} & \textbf{0.3594$\pm$0.1171} \\
\bottomrule
\end{tabular}%
}
\vspace{0.2mm}
\parbox{0.92\textwidth}{\scriptsize\emph{Note:} $\tau$ is calibrated per run at $\rho=0.80$; Known Reject is $1-\mathrm{TPR}(\tau)$ and Unknown FAR is $\mathrm{FPR}(\tau)$. MSP $W_1$ is the Wasserstein-1 distance between source-validation-known and target-known MSP distributions. For normalized target embeddings and class centroid $\mu_c^t$, Within-Class Cos. Dist. averages $1-\tilde z_i^\top\mu_{y_i}^t$. Unknown Intrusion is the fraction of unknown embeddings inside at least one class radius, where each radius is the 95th percentile of its target-known distances. Paired Wilcoxon tests give $p=0.0039$ for all diagnostics except Unknown FAR ($p=0.0078$).}
\end{table*}

The final full objective reduces target-known rejection by 12.72 percentage points relative to DSA alone. It also lowers deployment FAR by 0.69 percentage points, source-to-target MSP $W_1$ by 0.0141, target within-class cosine dispersion by 0.0085, and unknown intrusion by 32.65 percentage points. Each of these five paired comparisons yields $p<0.05$. Accordingly, the diagnostics support the scoped claim that the complete configuration stabilizes target-known confidence and reduces unknown penetration into known regions; they do not by themselves establish literal geometric shrinkage of the classifier boundary.

\begin{figure*}[!t]
\centering
\includegraphics[width=0.98\textwidth]{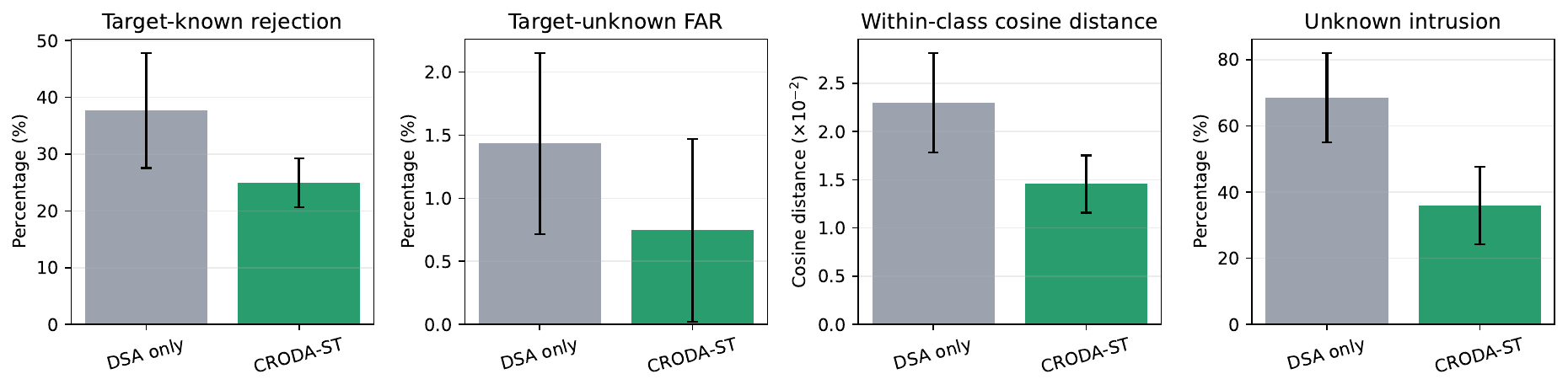}
\caption{Paired mechanism diagnostics for DSA-only and final full-objective checkpoints over three class splits and three seeds. Error bars indicate standard deviations. The source-calibrated operating point uses $\rho=0.80$ independently in each run.}
\label{fig:ablation-threshold-geometry}
\end{figure*}

\begin{figure*}[!t]
\centering
\includegraphics[width=0.84\textwidth]{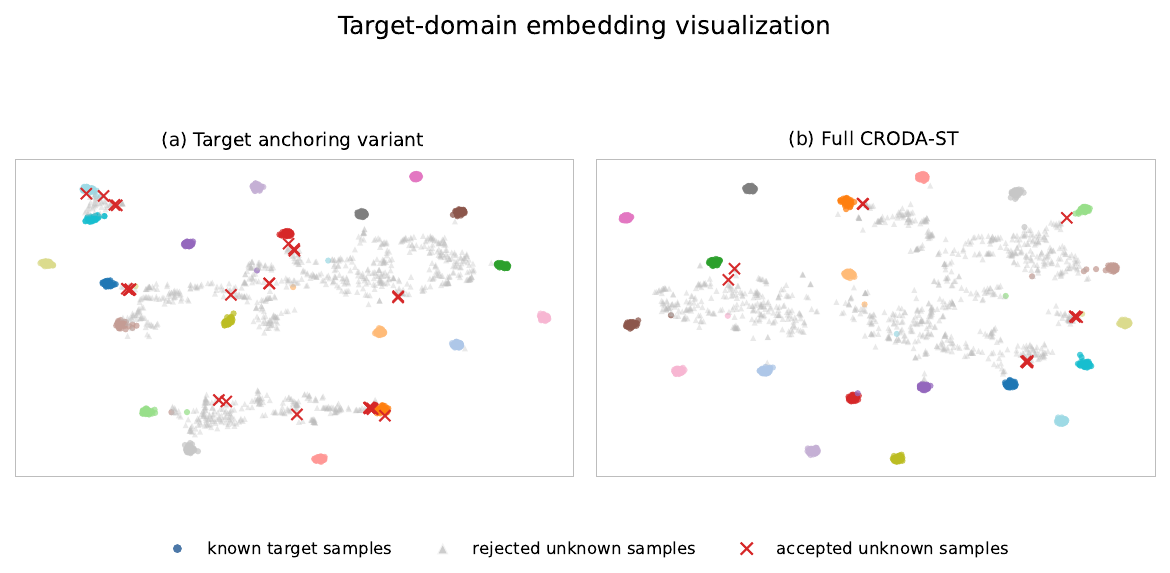}
\caption{Target-domain t-distributed stochastic neighbor embedding (t-SNE) visualization in the canonical 1-1$\rightarrow$1-19 setting.
Colored circles represent sampled known target classes, gray triangles indicate successfully rejected unknown samples, and red crosses denote false-positive accepted unknown samples.
CRODA-ST forms more compact known-class clusters and leaves fewer unknown samples in high-confidence known regions.}
\label{fig:target-embedding-tsne}
\end{figure*}

The target feature-space visualization shows a qualitative pattern consistent with the component analysis.
As illustrated in Fig.~\ref{fig:target-embedding-tsne}, the target-anchoring variant localizes target-known samples, while the full-objective visualization shows a more compact pattern alongside fewer false-positive unknown samples in high-confidence known regions.
The lower within-class cosine distance and unknown intrusion in Table~\ref{tab:st-mechanism-diagnostics} provide quantitative companion evidence.
Together, these qualitative and quantitative diagnostics suggest that the reduction in $\mathrm{FPR}_{90}$ is not attributable solely to score rescaling; they do not establish literal geometric shrinkage of the classifier boundary.

\subsection{Sensitivity to Warm-Up Order}
To examine whether joint optimization depends strongly on the order in which DSA and ROA are introduced, we compare two diagnostic schedules. One applies DSA alone during the first 40 epochs, whereas the other applies ROA alone; both then activate the full joint objective for the remaining 40 epochs.
Table~\ref{tab:st-warmup-robustness} reports the resulting run-averaged metrics.

\begin{table}[!t]
\caption{Sensitivity to DSA-first and ROA-first warm-up schedules before joint optimization.}
\label{tab:st-warmup-robustness}
\centering
\scriptsize
\setlength{\tabcolsep}{6pt}
\renewcommand{\arraystretch}{1.05}
\begin{tabular}{lccc}
\toprule
Setting & $\mathrm{Acc}_{K}$ $\uparrow$ & AUROC $\uparrow$ & OSCR $\uparrow$ \\
\midrule
DSA warm-up $\rightarrow$ full & \textbf{0.8595} & 0.9365 & 0.9120 \\
ROA warm-up $\rightarrow$ full & 0.8576 & \textbf{0.9440} & \textbf{0.9213} \\
\bottomrule
\end{tabular}
\vspace{0.8mm}
\parbox{0.95\columnwidth}{\scriptsize\emph{Note:} Results are averaged over 9 runs (3 splits $\times$ 3 seeds) with 40 warm-up epochs followed by 40 epochs of joint training. These schedules are diagnostic variants of the canonical training procedure.}
\end{table}

The ROA-first schedule yields nearly the same mean known-class accuracy as DSA-first (0.8576 versus 0.8595), while its mean AUROC and OSCR are higher by 0.0075 and 0.0093, respectively. Because this compact diagnostic reports run-averaged values without dispersion or paired significance tests, these differences are interpreted descriptively rather than as evidence that one order is superior. The experiment therefore indicates that neither tested warm-up order causes collapse and that the joint objective does not show a strong dependence on a strict DSA-first schedule under this setting.

\subsection{Temporal Shift and Supplementary Analyses}
\begin{table}[!t]
\caption{Target-domain adaptation and source-domain preservation under same-day and cross-day target data.}
\label{tab:st-cross-day-target}
\centering
\tiny
\setlength{\tabcolsep}{1.2pt}
\renewcommand{\arraystretch}{0.76}
\resizebox{0.98\columnwidth}{!}{%
\begin{tabular}{@{}llcccc@{}}
\toprule
Domain & L/U & \multicolumn{2}{c}{Same-day} & \multicolumn{2}{c}{Cross-day} \\
\cmidrule(lr){3-4}\cmidrule(lr){5-6}
 &  & $\mathrm{Acc}_{K}$ & OSCR & $\mathrm{Acc}_{K}$ & OSCR \\
\midrule
Target & 2/6 & 0.7017$\pm$0.0442 & 0.7753$\pm$0.0287 & 0.6772$\pm$0.0426 & 0.7415$\pm$0.0173 \\
Target & 3/5 & 0.7806$\pm$0.0262 & 0.8370$\pm$0.0205 & 0.7467$\pm$0.0423 & 0.7855$\pm$0.0203 \\
Target & 4/4 & 0.8153$\pm$0.0177 & 0.8489$\pm$0.0155 & 0.8125$\pm$0.0259 & 0.8284$\pm$0.0217 \\
\midrule
Source & 2/6 & 0.9441$\pm$0.0152 & 0.9492$\pm$0.0147 & 0.9405$\pm$0.0118 & 0.9465$\pm$0.0153 \\
Source & 3/5 & 0.9508$\pm$0.0126 & 0.9577$\pm$0.0137 & 0.9435$\pm$0.0109 & 0.9469$\pm$0.0139 \\
Source & 4/4 & 0.9479$\pm$0.0071 & 0.9490$\pm$0.0114 & 0.9443$\pm$0.0099 & 0.9481$\pm$0.0157 \\
\bottomrule
\end{tabular}%
}
\vspace{0.2mm}
\parbox{0.98\columnwidth}{\tiny\emph{Note:} L/U denotes labeled/unlabeled auxiliary samples from target known classes. This table reports a diagnostic variant rather than the main protocol.
In this variant, the U samples are used only as unlabeled target-domain data for receiver-discrimination and VAT regularization.
Their labels are not used for the target cross-entropy loss, supervised contrastive learning, threshold calibration, or model selection.
Source rows use the held-out source test split.}
\end{table}

Cross-day evaluation produces measurable but bounded degradation across the three few-shot settings in Table~\ref{tab:st-cross-day-target}. With 2/6 and 3/5 labeled/unlabeled auxiliary samples, target OSCR decreases by 0.0338 and 0.0515, respectively, and known-class accuracy decreases by 0.0245 and 0.0339. Under the 4/4 setting, the corresponding gaps narrow to 0.0205 in OSCR and 0.0028 in known-class accuracy, with cross-day accuracy reaching 0.8125 versus 0.8153 on the same day. Source-domain metrics vary by at most 0.0108 across the same comparisons. Thus, temporal shift is not eliminated, but the 4/4 setting preserves target recognition more closely while maintaining source-domain performance under this diagnostic protocol.

The supplementary analyses extend the evaluation across the tested receiver pairs and openness levels, with retention sensitivity and the controllable LoRa simulation reported alongside them in Appendix~\ref{app:supplementary-experiments}. The results provide breadth beyond the canonical pair and openness setting, but the elevated $\mathrm{FPR}_{90}$ values on difficult receiver pairs mark a remaining rejection boundary. They do not establish universal cross-receiver robustness.

For online inference, CRODA-ST remains an encoder--classifier path because the receiver discriminator, VAT perturbation, and supervised contrastive objective are used only during training.

\section{Conclusion}
CRODA-ST addresses the single-source single-target cross-receiver open-set RFFI challenge by recasting the recognition--rejection tradeoff as a threshold-reliability problem. Its joint, structure-anchored design couples explicitly anchored target-known spatial references with receiver-oriented confidence-boundary regularization. In this deployment setting, the target receiver must handle shifted known transmitters and unseen rogue devices under a threshold calibrated from source validation data. The experiments show that adaptation models can retain relatively high known-class accuracy while exhibiting high unknown acceptance at the ranking-based $\mathrm{FPR}_{90}$ operating point.

In the canonical WiSig ManyTx setting, CRODA-ST achieves a target known-class accuracy of 0.9092 and a target $\mathrm{FPR}_{90}$ of 0.0469. At the source-calibrated deployment operating point $\rho=0.80$, it yields a target-unknown deployment FAR of 0.0075$\pm$0.0072; this operating point is distinct from the ranking-based $\mathrm{FPR}_{90}$ operating point. Complementary evaluation on a controllable LoRa simulation provides a cross-protocol check under synthesized hardware distortions rather than empirical over-the-air validation. The matched-protocol component analysis and paired feature-geometry diagnostics provide configuration-level evidence consistent with complementary roles for structural anchoring and receiver-oriented regularization. Across the evaluated splits and seeds, the complete-configuration checkpoints exhibit lower target-known rejection, target within-class dispersion, and unknown intrusion than the DSA-only checkpoints, but these comparisons do not constitute strict one-factor mechanism proof.

Temporal, openness, and receiver-pair analyses extend this evidence across tested conditions beyond the canonical setting. Taken together, the results support structure-anchored cross-receiver RFFI under the evaluated receiver conditions. Performance on the hardest receiver pairs defines the current unknown-rejection boundary.

\clearpage
\appendix
\setcounter{table}{0}
\renewcommand{\thetable}{A\arabic{table}}
\makeatletter
\@ifundefined{theHtable}{}{\renewcommand{\theHtable}{A\arabic{table}}}
\makeatother

\section{Supplementary Experimental Results}
\label{app:supplementary-experiments}

\subsection{Sensitivity to Source-Calibrated Retention}
\label{app:retention-sensitivity}
\begin{table*}[!t]
\caption{Sensitivity to the retention rate $\rho$ in the canonical 1-1$\rightarrow$1-19 setting.}
\label{tab:st-rho-sensitivity}
\centering
\scriptsize
\setlength{\tabcolsep}{5pt}
\renewcommand{\arraystretch}{0.92}
\begin{tabular}{ccc}
\toprule
$\rho$ & $\tau$ & Unknown FAR@$\tau$ $\downarrow$ \\
\midrule
0.40 & 0.9984$\pm$0.0012 & \textbf{0.0013$\pm$0.0030} \\
0.45 & 0.9981$\pm$0.0013 & 0.0014$\pm$0.0032 \\
0.50 & 0.9979$\pm$0.0014 & 0.0015$\pm$0.0032 \\
0.55 & 0.9975$\pm$0.0015 & 0.0020$\pm$0.0033 \\
0.60 & 0.9971$\pm$0.0017 & 0.0025$\pm$0.0036 \\
0.65 & 0.9966$\pm$0.0020 & 0.0035$\pm$0.0046 \\
0.70 & 0.9959$\pm$0.0024 & 0.0044$\pm$0.0055 \\
0.75 & 0.9949$\pm$0.0028 & 0.0062$\pm$0.0065 \\
0.80 & 0.9936$\pm$0.0035 & 0.0075$\pm$0.0072 \\
0.85 & 0.9913$\pm$0.0046 & 0.0112$\pm$0.0087 \\
\bottomrule
\end{tabular}
\vspace{0.6mm}
\parbox{0.95\textwidth}{\scriptsize\emph{Note:} This table uses the same nine final full-objective checkpoints and canonical protocol as Table~\ref{tab:st-main-comparison-target}: receiver pair 1-1$\rightarrow$1-19, 37 known training classes among 114 target-test classes, 20 labeled target-enrollment samples per known class, and 9 runs from three class splits and three random seeds. For each run, $\tau$ is recalibrated only from source-validation known scores at retention rate $\rho$. Unknown FAR@$\tau$ equals $\mathrm{FPR}(\tau)$, the fraction of target-unknown samples accepted by this deployment threshold. AUROC, OSCR, and $\mathrm{FPR}_{90}$ are not repeated because they do not use $\tau$.}
\end{table*}

This analysis changes only the deployment threshold $\tau$ after training, while keeping the encoder, classifier, and knownness scores fixed. As $\rho$ increases from 0.40 to 0.85, the threshold becomes more permissive, decreasing from 0.9984$\pm$0.0012 to 0.9913$\pm$0.0046. The corresponding target-unknown FAR increases from 0.0013$\pm$0.0030 to 0.0112$\pm$0.0087. This confirms that $\rho$ controls the MSP operating point rather than the ranking quality of the knownness score itself.

In practical IoT deployments, the selection of $\rho$ represents a security--usability tradeoff, although Table~\ref{tab:st-rho-sensitivity} reports target-unknown FAR rather than a direct usability measure. Because $\rho$ is set on the source validation set $\mathcal{D}_{\mathrm{val}}^s$, no target-domain unknowns are required. The tested operating points illustrate how $\rho$ can be associated with different application priorities:
\begin{itemize}
    \item \textbf{High-Security Scenarios:} For deployments prioritizing strict access control (e.g., critical infrastructure), the lower tested range (e.g., $0.40 - 0.60$) is associated with a stricter source-calibrated operating point. Across the tested $\rho=0.40$--$0.60$ range, the mean target-unknown deployment FAR remains below 0.003.
    \item \textbf{High-Usability Scenarios:} For environments where registered device availability is prioritized (e.g., smart home access), a larger $\rho$ gives a more permissive operating point, with a correspondingly higher but still bounded unknown-acceptance risk.
\end{itemize}
These ranges are illustrative for the tested protocol rather than universal administrative prescriptions.

\subsection{Openness Sensitivity}
\label{app:openness-sensitivity}
\begin{table}[!t]
\caption{Sensitivity to openness in the canonical receiver pair 1-1$\rightarrow$1-19.}
\label{tab:st-openness-sensitivity}
\centering
\tiny
\renewcommand{\arraystretch}{0.76}
\resizebox{0.88\columnwidth}{!}{%
\begin{tabular}{ccccc}
\toprule
Open. & $K_{\rm tr}$/$K_{\rm te}$ & $\mathrm{Acc}_{K}$ $\uparrow$ & OSCR $\uparrow$ & $\mathrm{FPR}_{90}$ $\downarrow$ \\
\midrule
0.10 & 81/119 & 0.9051$\pm$0.0140 & 0.9318$\pm$0.0104 & 0.1138$\pm$0.0413 \\
0.19 & 64/133 & 0.9193$\pm$0.0159 & 0.9467$\pm$0.0171 & 0.0764$\pm$0.0397 \\
0.30 & 37/114 & 0.9092$\pm$0.0182 & \textbf{0.9580$\pm$0.0122} & \textbf{0.0469$\pm$0.0309} \\
0.40 & 27/123 & 0.8793$\pm$0.0215 & 0.9244$\pm$0.0168 & 0.1640$\pm$0.0697 \\
0.50 & 19/133 & 0.9275$\pm$0.0247 & 0.9520$\pm$0.0173 & 0.0957$\pm$0.0403 \\
0.59 & 12/133 & \textbf{0.9364$\pm$0.0406} & 0.9386$\pm$0.0450 & 0.1391$\pm$0.1070 \\
\bottomrule
\end{tabular}%
}
\vspace{0.2mm}
\parbox{0.88\columnwidth}{\tiny\emph{Note:} $K_{\rm tr}/K_{\rm te}$ denotes known training classes / total test classes, and openness is $\mathcal{O}=1-\sqrt{2K_{\rm tr}/(K_{\rm tr}+K_{\rm te})}$. Each openness level uses 20 labeled target-enrollment samples per known class and is averaged over three class splits and three random seeds. The openness 0.30 row is the canonical setting reported in Table~\ref{tab:st-main-comparison-target}.}
\end{table}

Across the tested openness levels, $\mathrm{FPR}_{90}$ follows a nonmonotonic pattern: it is 0.1138 at openness 0.10, 0.0469 at 0.30, 0.1640 at 0.40, and 0.0957 at 0.50. Openness alone therefore does not form a monotonic rejection trend in this table. Because both the class identities and the $K_{\rm tr}/K_{\rm te}$ composition vary across rows, the table cannot establish hardware-fingerprint overlap as the cause of this pattern.

\subsection{Online Inference Cost}
\label{app:online-inference-cost}
\begin{table}[!t]
\caption{Online inference cost of CRODA-ST in the canonical cross-receiver setting.}
\label{tab:st-inference-cost}
\centering
\tiny
\renewcommand{\arraystretch}{0.84}
\resizebox{0.92\columnwidth}{!}{%
\begin{tabular}{lc}
\toprule
Quantity & Value \\
\midrule
Trainable parameters & 0.262M \\
32-bit floating-point (FP32) model storage & 1.05 MB \\
MACs per $2\times256$ I/Q sample & 5.29M \\
FLOPs per sample & 10.48M \\
Single-thread CPU latency & 1.35 ms/sample \\
Batch-64 amortized CPU latency & 0.51 ms/sample \\
\bottomrule
\end{tabular}%
}
\vspace{0.2mm}
\parbox{0.92\columnwidth}{\tiny\emph{Note:} The FLOP count treats one multiply--accumulate as two floating-point operations. CPU latency is measured over 800 forward passes on an Intel Core i9-12900HX under the 37-known-class canonical setting.}
\end{table}

At deployment, CRODA-ST passes each I/Q segment once through the 2-D residual encoder and the linear classifier. It then performs open-set rejection by comparing the maximum softmax probability with the deployment threshold. The receiver discriminator, VAT perturbation, and supervised contrastive objective serve only as training-time constraints and are removed from online authentication. Therefore, CRODA-ST introduces no additional online model branch beyond the standard encoder--classifier path.

\subsection{Receiver-Pair Robustness}
\label{app:receiver-pair-robustness}
\begin{table}[!t]
\caption{Robustness on 10 randomly selected source-target receiver pairs.}
\label{tab:st-pair-robustness}
\centering
\tiny
\renewcommand{\arraystretch}{0.76}
\resizebox{0.84\columnwidth}{!}{%
\begin{tabular}{lccc}
\toprule
Pair & $\mathrm{Acc}_{K}$ $\uparrow$ & OSCR $\uparrow$ & $\mathrm{FPR}_{90}$ $\downarrow$ \\
\midrule
1-1$\rightarrow$14-7 & 0.9109 & 0.9383 & 0.0952 \\
7-14$\rightarrow$19-2 & 0.8907 & 0.9153 & 0.1419 \\
1-1$\rightarrow$7-7 & 0.9087 & 0.9406 & 0.1029 \\
14-7$\rightarrow$1-1 & 0.9502 & 0.9567 & 0.0642 \\
1-1$\rightarrow$1-19$^{\dagger}$ & 0.9092 & 0.9580 & 0.0469 \\
20-19$\rightarrow$19-1 & 0.9367 & 0.9463 & 0.0736 \\
14-7$\rightarrow$19-1 & 0.9285 & 0.9336 & 0.1182 \\
14-7$\rightarrow$20-1 & 0.8739 & 0.8867 & 0.2354 \\
1-1$\rightarrow$20-1 & 0.8565 & 0.8641 & 0.2815 \\
7-7$\rightarrow$20-19 & 0.9251 & 0.9331 & 0.1246 \\
\bottomrule
\end{tabular}%
}
\vspace{0.2mm}
\parbox{0.84\columnwidth}{\tiny\emph{Note:} All rows use the openness-0.30 class split protocol with 37 known training classes among 114 target-test classes and 20 labeled target-enrollment samples per known class. Each non-canonical pair is averaged over 9 runs from three class splits and three random seeds. $\dagger$ uses the canonical 1-1$\rightarrow$1-19 result in Table~\ref{tab:st-main-comparison-target}; the table reports pair-level means and is not a 10-pair average. AUROC is omitted from this compact pair-summary table. OSCR: open-set classification rate; $\mathrm{FPR}_{90}$: false positive rate at 90\% true positive rate, distinct from the deployment threshold $\tau$.}
\end{table}

Across these 10 selected configurations, known-class accuracy ranges from 0.8565 to 0.9502, whereas $\mathrm{FPR}_{90}$ exhibits larger variation. The two most challenging receiver pairs reach $\mathrm{FPR}_{90}$ values of 0.2354 and 0.2815. Unknown rejection therefore remains the primary residual difficulty for these selected configurations. The table does not directly measure hardware-distortion severity or known--unknown margin compression and therefore does not attribute the pair-level differences to either factor.

\subsection{Controllable LoRa Simulation Array}
\label{app:lora-array}
\label{sec:exp-lora-array}
To provide a complementary simulation-based check beyond the WiFi dataset, we evaluate CRODA-ST on a synthetic LoRa protocol with ten systematically controlled receiver hardware configurations ($\mathrm{LoRa\text{-}1}$ to $\mathrm{LoRa\text{-}10}$).
The simulation injects receiver-dependent impairments, including carrier-frequency offset, sampling-frequency offset, I/Q imbalance, phase offset, front-end filtering variation, and additive noise.
These controlled factors are intended to probe whether the open-set formulation remains usable under synthesized hardware distortions, rather than to replace empirical over-the-air validation.
Table~\ref{tab:st-lora-top5} reports selected cross-receiver transfer pairs across both target and source test domains.

\begin{table}[!t]
\caption{Target- and source-domain open-set performance on selected LoRa receiver pairs.}
\label{tab:st-lora-top5}
\centering
\tiny
\renewcommand{\arraystretch}{0.80}
\resizebox{0.96\columnwidth}{!}{%
\begin{tabular}{lcccc}
\toprule
Receiver Pair & $\mathrm{Acc}_{K}$ $\uparrow$ & AUROC $\uparrow$ & OSCR $\uparrow$ & $\mathrm{FPR}_{90}$ $\downarrow$ \\
\midrule
\multicolumn{5}{l}{\textit{Target-domain test}} \\
LoRa-5 $\rightarrow$ LoRa-1 & 0.9620 & 0.9452 & 0.9452 & 0.0962 \\
LoRa-7 $\rightarrow$ LoRa-3 & 0.9780 & 0.9416 & 0.9416 & 0.1514 \\
LoRa-9 $\rightarrow$ LoRa-4 & 0.9840 & 0.9310 & 0.9310 & 0.1090 \\
LoRa-4 $\rightarrow$ LoRa-6 & 0.9940 & 0.9226 & 0.9226 & 0.1600 \\
LoRa-9 $\rightarrow$ LoRa-6 & 1.0000 & 0.9225 & 0.9225 & 0.1295 \\
\textit{Target Mean} & \textbf{0.9836} & \textbf{0.9326} & \textbf{0.9326} & \textbf{0.1292} \\
\midrule
\multicolumn{5}{l}{\textit{Source-domain test (Preservation)}} \\
LoRa-5 $\rightarrow$ LoRa-1 & 0.9950 & 0.9860 & 0.9855 & 0.0285 \\
LoRa-7 $\rightarrow$ LoRa-3 & 0.9960 & 0.9812 & 0.9810 & 0.0310 \\
LoRa-9 $\rightarrow$ LoRa-4 & 0.9980 & 0.9895 & 0.9890 & 0.0210 \\
LoRa-4 $\rightarrow$ LoRa-6 & 0.9970 & 0.9840 & 0.9835 & 0.0260 \\
LoRa-9 $\rightarrow$ LoRa-6 & 1.0000 & 0.9910 & 0.9910 & 0.0180 \\
\textit{Source Mean} & \textbf{0.9972} & \textbf{0.9863} & \textbf{0.9860} & \textbf{0.0249} \\
\bottomrule
\end{tabular}%
}
\vspace{0.2mm}
\parbox{0.96\columnwidth}{\tiny\emph{Note:} Results follow the same protocol. $\mathrm{FPR}_{90}$ is the false positive rate at the threshold that retains 90\% of known samples in the evaluated domain. Target rows evaluate open-set classification on the target receiver, while source rows confirm identity preservation on the original source receiver.}
\end{table}

Across the five selected target receiver pairs, CRODA-ST achieves a target mean known-class accuracy of 0.9836 and a target mean $\mathrm{FPR}_{90}$ of 0.1292. The corresponding source mean known-class accuracy is 0.9972. These measurements provide a controlled cross-protocol check under synthesized hardware impairments; they do not replace empirical cross-protocol over-the-air validation.

\printcredits

\bibliographystyle{cas-model2-names}
\bibliography{refs_st_en}

\end{document}